\documentclass[twoside,12pt]{article}
\usepackage{epsfig}

\def\Journal#1#2#3#4{{#1} {#2} (#4) #3}

\def\NPB{{\em Nucl. Phys.} B}

\newcommand{\be}{\begin{equation}}
\newcommand{\ee}{\end{equation}}
\newcommand{\bea}{\begin{eqnarray}}
\newcommand{\eea}{\end{eqnarray}}

\begin{document}

\title{ \vspace{1cm} Effect of quark off-shellness in DIS and the Drell-Yan process}
\author{O.\ Linnyk, S.\ Leupold, U.\ Mosel\\
\\
Institut f\"ur Theoretische Physik, Universit\"at Giessen, Germany}
\maketitle
\begin{abstract} 
We study higher twist corrections to the perturbative QCD cross sections for
$ep$ deep inelastic scattering and the Drell-Yan lepton pair production in
$pp$ collision. The corrections arise due to the initial state interaction of
the active quarks with the spectator partons. The effect of this
interaction is calculated by dressing the incoming quark lines with phenomenological single
parameter Breit-Wigner spectral functions and taking into account full off-shell kinematics. 
The quark width in the proton is estimated by analyzing the data on Drell-Yan
triple differential cross section from the experiment E866 at Fermilab.
\end{abstract}
%\eject
%\tableofcontents
%
%\section{Introduction and Method  \label{intro}}
%
There is an analogy between the quasi-elastic $e$-nucleus scattering at
high energy and the deep inelastic $ep$ scattering(DIS).
In the plane wave impulse approximation(PWIA) of nuclear physics, the
electron is believed to interact with single nucleon in the
nucleus, if the energy transfer is much higher than the binding energy. 
Inclusive $eA$ cross sections can be calculated from a 
handbag diagram, neglecting the interaction of the struck nucleon with the rest
of the nucleus. 
In this approximation, the nucleon is treated as free, and the on-shell relation fixes
one component of the nucleon momentum: $\! p^2\! =\! M_N^2$. 
%It was shown in many works (for instance,~\cite{nonPWIA}) that 
However,
one has to go beyond PWIA in order to
describe (semi-)exclusive observables, such as the nucleon knock-out reactions.
In particular, one has to take into account 
the nucleon off-shellness due to the initial state interaction (ISI).
\vspace{-0.5cm}
\begin{figure}[tb]  
\begin{minipage}[b]{0.45\linewidth} % A minipage that covers half the page, width-wise  
\begin{center}
\epsfig{file=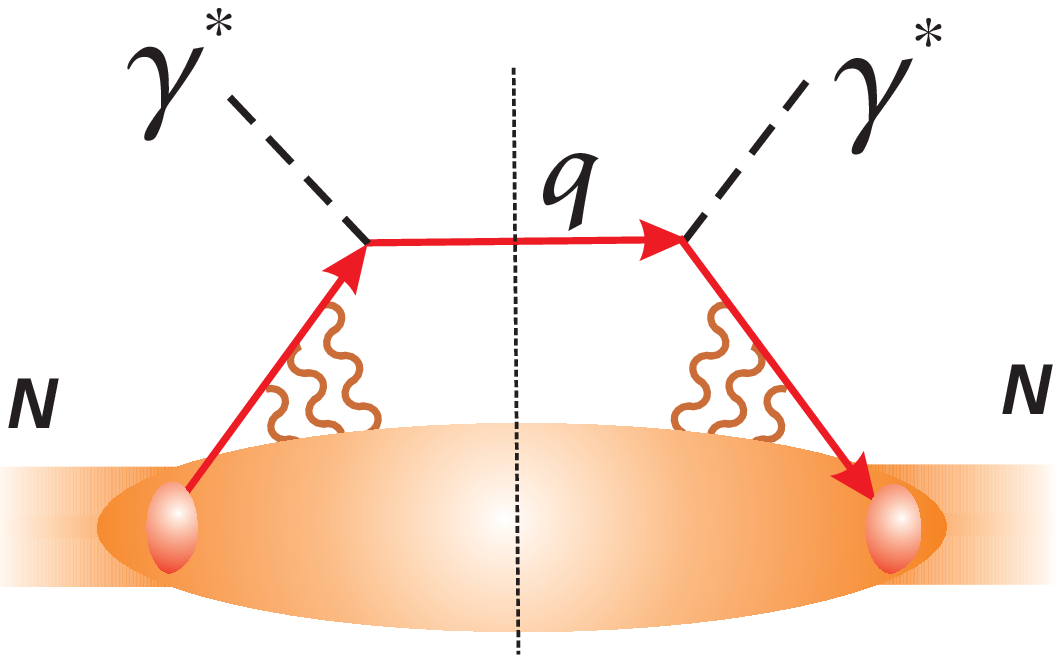,scale=0.7}  
\caption{The handbag graph for DIS and the relevant initial 
state interactions that could build up a finite parton width \label{diagram}}  
\end{center}
\end{minipage}  
\hspace{0.5cm} % To get a little bit of space between the figures  
\begin{minipage}[b]{0.5\linewidth}  
\centering \epsfig{file=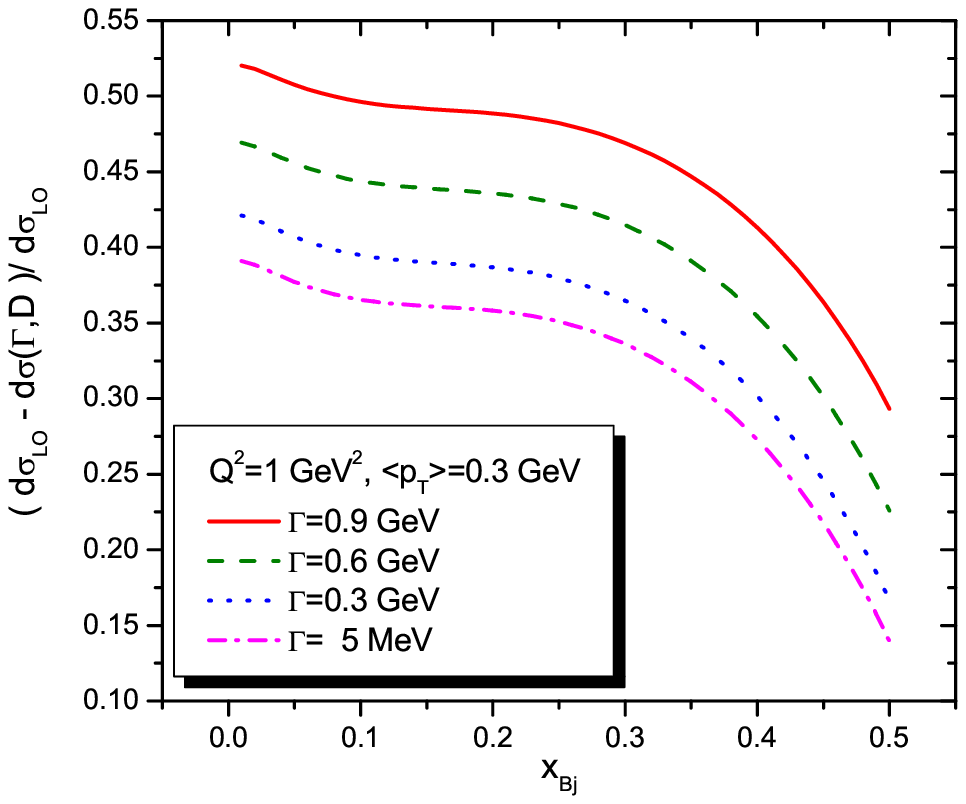,scale=0.65}  
\caption{Deviation of the calculated DIS cross section from parton model
for a range of parton spectral function widths.\label{DIS}}  
\end{minipage} 
\end{figure}

In case of fully inclusive $ep$ DIS, the factorization theorem justifies the use of
the handbag diagram at high momentum transfer squared $Q^2$. 
In the limit $\! Q^2 \! \to \! \infty  $, the quark can be considered free and its
momentum satisfies the on-shell relation $ p^- \! = \! p^0 \! - \! p^z
\! = \! 0 $. 
On the other hand, a quark bound in nucleon is always off-shell, {\it i.e.} $p^-$ is not
fixed to $0$, but is distributed with some finite width.
At moderate $Q^2$ the effect of quark off-shellness on the observable cross
section might be considerable.
The handbag diagram for the DIS with the relevant initial state interactions
is presented in Fig.~\ref{diagram}.
Following the suggestion of~\cite{benhar.N}, we calculated the quark
off-shellness effects by dressing the incoming quark lines with model spectral functions. 
We use the factorization assumption to write the DIS cross section as follows: 
\be 
\label{g_fact_DIS}
d \sigma _{\mbox{\small DIS}} = 
\sum _i
g_i ( \xi, \vec p_{\perp} , p^-) \otimes 
d \hat \sigma_i (\xi, \vec p_{\perp} ,p^-),
\ee 
where $d \hat \sigma_i$ is the cross section of the sub-process $eq_i\to eq_i$,
$\xi=p^+/P_N^+$ is the fraction of nucleon light cone momentum carried by the
struck quark, $p^-$ is the minus component of the quark's momentum and $\vec
p_{\perp}$ is its transverse momentum. 
$d \hat \sigma_i$  is convoluted with a function 
$g_i ( \xi, \vec p_{\perp} , p^-)$ giving the probability to find 
a quark of flavour $i$ with momentum $p$ and virtuality (off-shellness) 
$m^2$ in the nucleon ($m^2 \equiv p^+p^- -\vec p_{\perp}^2 $). 
Formula (\ref{g_fact_DIS}) is easily generalized for the case of two partons 
in the initial state to describe the Drell-Yan process. In addition, we propose
the following ansatz:
\be
g_i ( \xi, \vec p_{\perp} , p^-) =
f_i ( \xi, \vec p_{\perp}) A(m,\Gamma),
\ee
where $f_i ( \xi, \vec p_{\perp})$ are unintegrated parton
distributions~\cite{mulders}, $A(m,\Gamma)$ quark spectral function, for which
we use the Breit-Wigner parametrization with constant width.
%
%\section{Results}
%
\begin{figure}[tb]  
\begin{minipage}[b]{0.45\linewidth} % A minipage that covers half the page, width-wise  
\begin{center}
\epsfig{file=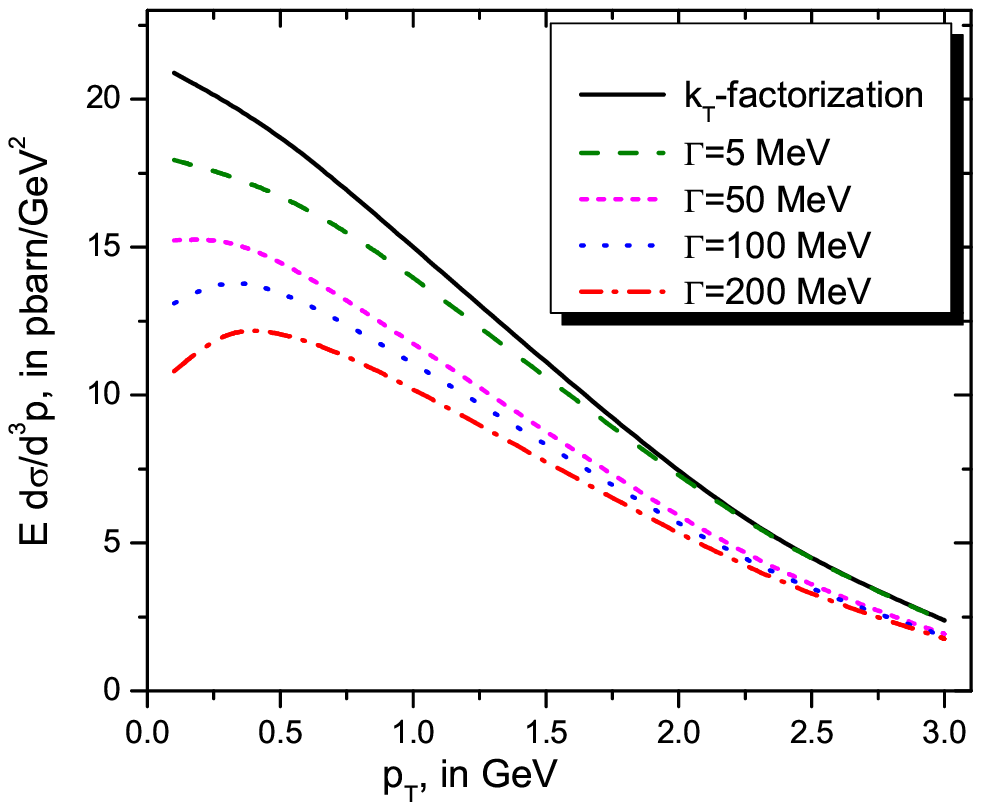,scale=0.7}  
\caption{Triple differential Drell-Yan cross section in our model for different values of 
the parton width~$\Gamma$. The solid line is at $\Gamma=0$.\label{DY1}}  
\end{center}
\end{minipage}  
\hspace{0.5cm} % To get a little bit of space between the figures  
\begin{minipage}[b]{0.5\linewidth}  
\centering \epsfig{file=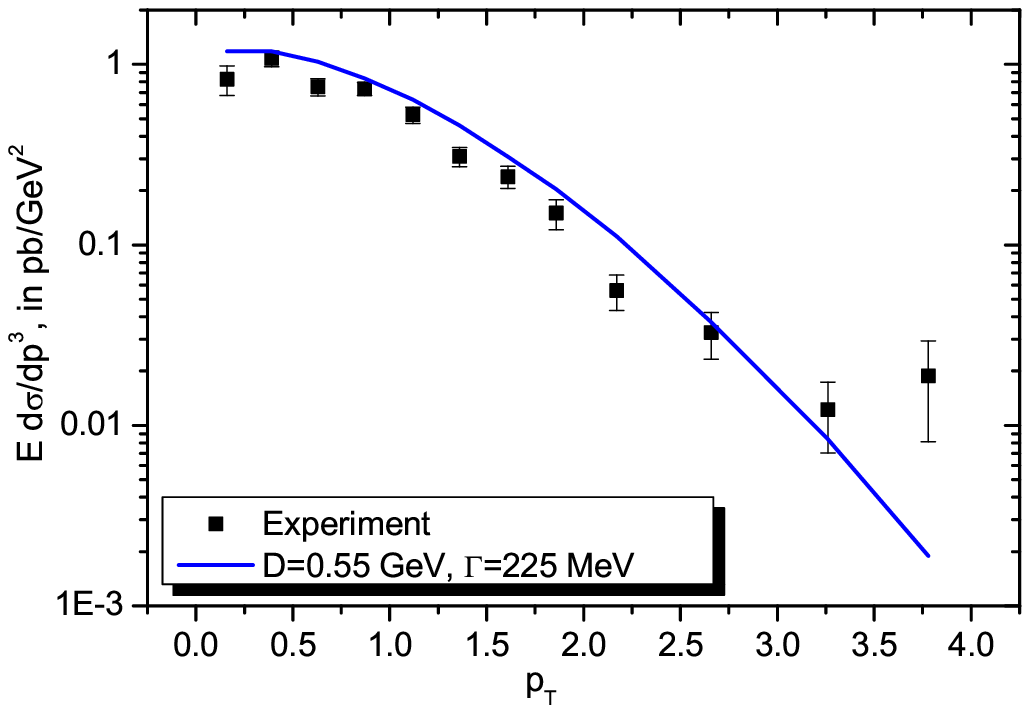,scale=0.78}  
\caption{Calculated Drell-Yan cross section compared to the data of 
the Fermilab experiment 866 for $pp\to \mu^+\mu^-+X$ at 800~GeV
incident $p$ energy.\label{DY2}}  
\end{minipage} 
\end{figure}
%
%\section{Conclusion}

The results of our calculations for fully inclusive DIS are presented in
Figure~\ref{DIS}. 
The calculated Drell-Yan cross section for different values of $\Gamma$ are given in
Figure~\ref{DY1}. 
From the comparison to the data if the Fermilab experiment E866~\cite{exp}, we
estimated the quark width in proton to be of the order of 200 MeV. 
The example of our description of the data is presented in Figure~\ref{DY2} in
one bin of muon pair mass $7.2\!\le\! M\!\le\!8.7$~GeV and Feynman x $-0.05\!\le\! x_F\!\le\!0.15$. 
The data can be described without a K-factor.

\end{document}